\documentclass[letterpaper,preprintnumbers,prd,nofootinbib,nobibnotes,showpacs]{revtex4}
\usepackage{amsfonts}
\usepackage{mathrsfs}
\usepackage{epsfig}
\usepackage{graphicx}%
\usepackage{dcolumn}
\usepackage{amsmath}
\usepackage{color}
\usepackage{color}
\makeatletter
\def\btt#1{\texttt{\@backslashchar#1}}%
\DeclareRobustCommand\bblash{\btt{\@backslashchar}}%
\makeatother
\begin{document}

\title{Gravity, matters and dark energy on the minimal surfaces}
\author{Changjun Gao}\email{gaocj@bao.ac.cn}\affiliation{ Key Laboratory of Computational Astrophysics, National Astronomical Observatories, Chinese Academy of Sciences, Beijing 100101, China}
\affiliation{School of Astronomy and Space Sciences, University of Chinese Academy of Sciences,
No. 19A, Yuquan Road, Beijing 100049, China}\author{Shuang Yu}\email{yushuang@nao.cas.cn} \affiliation{ Key Laboratory of Computational Astrophysics, National Astronomical Observatories, Chinese
Academy of Sciences, Beijing 100101, China}
\affiliation{School of Astronomy and Space Sciences, University of Chinese Academy of Sciences,
No. 19A, Yuquan Road, Beijing 100049, China}

\date{\today}

\begin{abstract}
We propose gravity, matters and dark energy may be confined on different four dimensional \emph{minimal surfaces} for the observer in five dimensional spacetime.
Following this idea, we construct the equations of motion when gravity, matters and dark energy are confined on different minimal surfaces. Then in the background of Friedman-Robertson-Walker universe, we investigate the cosmic evolution when gravity and quintessence are confined on the minimal surfaces.

\end{abstract}



\maketitle

\section{Introduction}
The acceleration of the universe is usually interpreted with the help of dark energy. Dark energy
can result from the cosmological constant, the ideal fluid, the manifestation of vacuum effects and
the scalar field etc (For a very complete review of dynamical dark energy see \cite{cope:2006} and
references therein). The majority of dark energy candidates reflects the undisputable fact that the true nature and origin of dark energy
has not been convincingly explained yet. What is more, there is another quite appealing possibility that dark energy is originated
from the modification of General Relativity. In practice, there are many classes of modified gravity models which have been widely
investigated such as $f(R)$ gravity models \cite{capo:2003} and the brane-world models \cite{DGP:2000}. In this paper we are particularly interested in brane world models.

The brane world models have been investigated in many phenomenological, cosmological and astrophysical aspects.  One of the remarkable conclusions of these models is that the size of extra dimensions can be very large \cite{ark:1998} or even infinite \cite{randall1:1999,gog:1998,greg:2000,randall2:1999}. This leads to many specific predictions that can be tested observationally now and in the coming future.
There are mainly two kinds of brane world scenarios. They are Randall-Sundrum models \cite{randall1:1999} (RS) and the Dvali-Gabadadze-Porrati \cite{DGP:2000} (DGP) model. The single-brane RS models with infinite extra dimension arise when the orbifold radius tends to infinity. This extra
dimension is curved or ¡°warped¡± rather than flat. The RS models
modify general relativity at high energies. By contrast, the DGP model modifies general relativity at low energies. Same as the RS model, the DGP model is a 5-dimensional model with infinite extra dimensions. Different from the Anti-de Sitter bulk of the RS model, the bulk in DGP is assumed to be 5-dimensional Minkowski spacetime.

Motivated by the idea of brane world, we consider the scenarios of gravity and matters confined on four dimensional \emph{minimal surfaces}. As is well known \emph{minimal surfaces} (or sub-manifold), most familiar as soap films, have great relevance to a variety of physical problems such as the positive energy conjecture, the materials science and civil engineering. In view of this point, we explore the relevance of minimal surfaces with gravity, matters and dark energy. To this end, one consider a five dimensional spacetime with the metric
\begin{equation}
ds^2=g_{\mu\nu}dx^{\mu}dx^{\nu}+h\left(x^{\alpha}\right)df^2\;,\nonumber\\
\end{equation}
where $\tilde{g}_{\mu\nu}(x^{\alpha})$ and $g_{\mu\nu}(x^{\alpha})$ are the metric of curved four dimensional spacetime and $f$ is the extra dimension. In this form, the extra dimension is curved due to the presence of $h(x^{\alpha})$. In the background of this bulk metric, we shall construct the dynamical equations
when gravity, matters and dark energy are confined on minimal surfaces. As simple examples, we investigate the evolution of the universe when gravity and quintessence are confined on the minimal surfaces, respectively.

The paper is organized as follows. In Sec. II, we derive
the equations of motion when gravity, matters and dark energy on the minimal surfaces. In Sec. III, in the background of Friedman-Robertson-Walker universe,
we present the corresponding dynamical equations. In Sec. IV, we consider the cosmic evolution when gravity is confined a minimal surface.  In Sec. V, we consider the cosmic evolution of quintessence when it is confined a minimal surface. Section VI gives the conclusion and discussion. Throughout this paper, we adopt the system of units in which $G=c=\hbar=1$ and the metric signature
$(-,\ +,\ +,\ +)$.

\section{equations of motion}
The string theory reveals that the dimension of our spacetime must be higher than four on small scales \cite{ruba:1983}. So the observational and macroscopic four dimensional universe must be a super-surface, for example  $f$ in Fig.~(1),  embedded in five dimensional spacetime. Here the metric of the 5-dimensional spacetime is assumed to be
\begin{equation}
ds^2=g_{MN}dx^{M}dx^{N}=g_{\mu\nu}\left(x^{\alpha}\right)dx^{\mu}dx^{\nu}+h\left(x^{\alpha}\right)df^2\;.
\end{equation}
$f$ is the extra fifth dimension and $x^{\alpha}=(t,x,y,z)$ are the coordinates of the four dimensional spacetime. We have ${x^M}=(x^{\alpha},f)$. The geometry of surface $f=f(x^{\alpha})$ is described by the metric $\tilde{g}_{\mu\nu}(x^{\mu})$ which is induced from the bulk 5-dimensional metric as follows
\begin{equation}
\tilde{g}_{\mu\nu}=g_{MN}\frac{\partial x^M}{\partial x^{\mu}}\frac{\partial x^N}{\partial x^{\mu}}=g_{\mu\nu}+h\frac{\partial f}{\partial x^{\mu}}\cdot\frac{\partial f}{\partial x^{\nu}}\;.
\end{equation}
We then have
\begin{equation}
\sqrt{-\tilde{g}}=\sqrt{-g}\cdot\sqrt{1+h\left(\nabla f\right)^2}\;,
\end{equation}
with
\begin{eqnarray}
\left(\nabla f\right)^2&\equiv&\nabla_{\mu}f\nabla^{\mu}f=g^{\mu\nu}\nabla_{\mu}f\nabla_{\nu}f\;.
\end{eqnarray}
Here  $\tilde{g}$ and $g$ are the determinant of four dimensional metrics  $\tilde{g}_{\mu\nu}$ and $g_{\mu\nu}$ on surfaces $f\neq const$ and $f=const$, respectively.  $\nabla_{\mu}$ is the corresponding four dimensional covariant derivative on $f=const$.

Then the area of the surface is given by
\begin{eqnarray}
A&=&\int\sqrt{1+h\left(\nabla f\right)^2}\sqrt{-g}d^4x\;.
\end{eqnarray}
When $h=const$ and $g_{\mu\nu}=\delta_{\mu}^{\nu}$, it is exactly the form of the well-known formulae in Euclidean space. However, here it is applicable to Riemann space.

Variation of the area with respect to $f$ gives the equation for minimal surface
\begin{eqnarray}
\nabla_{\mu}\left[\frac{h\nabla^{\mu}f}{\sqrt{1+h\left(\nabla f\right)^2}}\right]=0\;.
\end{eqnarray}
When $h=const$ and $g_{\mu\nu}=\delta_{\mu}^{\nu}$, it reduces to the famous equation for minimal surface in Euclidean space.
 It can be understood as the equation of motion for the ``scalar field" $f$. So, if the surface $f$ is regarded as a scalar field, the formulae for the  area can be understood as the action of ``scalar field'' $f$.
\begin{figure}[h]
\includegraphics[width=8cm]{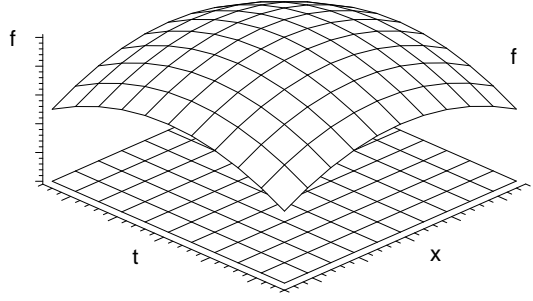}
\caption{Super-surface $f=f(x^{\mu})$ in the 5-dimensional spacetime.}
\end{figure}
In particular, when energy is distributed on the surface $f$ with density $\sigma$, the total energy on the surface is
\begin{eqnarray}
E&=&\int\sigma\sqrt{1+h\left(\nabla f\right)^2}\sqrt{-g}d^4x\;,
\end{eqnarray}
which is a well-known result.
\begin{figure}[h]
\includegraphics[width=8cm]{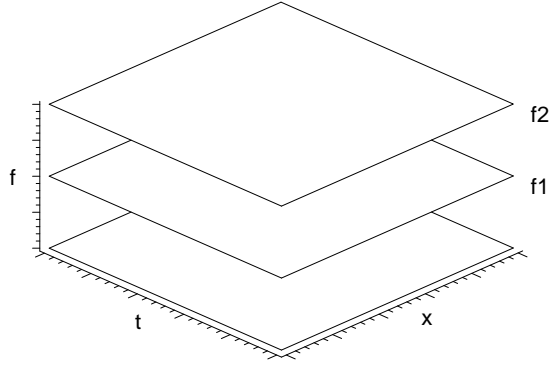}
\caption{Two surfaces, $f_1=f_1(x^{\mu})$ and $f_2=f_2(x^{\mu})$ in 5-dimensional spacetime.}
\end{figure}

Now we consider two surfaces, $f_1, f_2$ as shown in Fig.~(2). They are described by
\begin{eqnarray}
f_1=f_1\left(x^{\mu}\right)\;,\ \ \ \ f_2=f_2\left(x^{\mu}\right)\;,
\end{eqnarray}
respectively. Then the Einstein-Hilbert action on $f_1$ takes the form
\begin{eqnarray}
S_1=\int \frac{R}{16\pi}\sqrt{1+h\left(\nabla f_1\right)^2}\sqrt{-g}d^4x\;.
\end{eqnarray}
When $f_1=const$, it reduces to the usual form. We note that one can generally replace $R$ with some function of $K\left(R, R_{\mu\nu}R^{\mu\nu},R_{\mu\nu\alpha}^{\lambda}R^{\mu\nu\alpha}_{\lambda}\right)$ for modified gravity. But for simplicity in this study, we shall consider $K=R$.

On the other hand, if matters are confined on the surface $f_2$, the action for matters takes the form
\begin{eqnarray}
S_2=\int\sum_{i}\mathscr{L}^{i}\sqrt{1+h\left(\nabla f_2\right)^2}\sqrt{-g}d^4x\;,
\end{eqnarray}
where $\mathscr{L}^{i}$ is the Lagrangian of i-th component of matters.
Combining these considerations, we write the total action
\begin{eqnarray}
S=\int\left[\frac{R}{16\pi}\sqrt{1+h\left(\nabla f_1\right)^2}+\sum_{i}\mathscr{L}^{i}\sqrt{1+h\left(\nabla f_2\right)^2}\right]\sqrt{-g}d^4x\;.
\end{eqnarray}
When $f_1$ and $f_2$ are identified with a constant surface, the action reduces to the Einstein gravity. But in general it differs from the latter when $f_1\neq const$ and $f_2\neq const$. When $f_1$ is identified with $f_2$, both gravity and the matters are confined on the same surface. Of course, one can assume different matters are confined on different surfaces. But here we only consider the case of two surfaces.

Variation of the action with respect to $g^{\mu\nu}$ gives the Einstein equations
\begin{eqnarray}\label{form:1}
&&G_{\mu\nu}\sqrt{1+h\left(\nabla f_1\right)^2}+\frac{ Rh\nabla_{\mu}f_1\nabla_{\nu}f_1}{2\sqrt{1+h\left(\nabla f_1\right)^2}}-\left(\nabla_{\mu}\nabla_{\nu}-g_{\mu\nu}\nabla^2\right)\sqrt{1+h\left(\nabla f_1\right)^2}\nonumber\\&&=8\pi\left[\sum_{i}\bar{T}^{i}_{\mu\nu}\sqrt{1+h\left(\nabla f_2\right)^2}-\sum_{i}\mathscr{L}^{i}\frac{h\nabla_{\mu}f_2\nabla_{\nu}f_2}{\sqrt{1+h\left(\nabla f_2\right)^2}}\right]\;,
\end{eqnarray}
where
\begin{eqnarray}\label{form:2}
\bar{T}^{i}_{\mu\nu}\equiv-\frac{2}{\sqrt{-g}}\cdot\frac{\delta \left(\sqrt{-g}\mathscr{L}^{i}\right)}{\delta g^{\mu\nu}}\;,
\end{eqnarray}
is the energy-momentum tensor for i-th component of matters. It is well-known that the definition of matter Lagrangian giving the perfect
fluid energy-momentum tensor is not unique; one can choose either
$\mathscr{L}^{i}=p^{i}$ or  $\mathscr{L}^{i}=-\rho^{i}$ which provide the same energy-momentum tensor (see
\cite{bert:2008,far:2009} for a detailed discussion). Here $\rho^{i}$ and $p^{i}$ are the energy density and pressure, respectively. In the present study,
we shall consider $\mathscr{L}^{i}=-\rho^{i}$.

Variation of the action with respect to $f_i$ give the equations of motion for the minimal surfaces,
\begin{eqnarray}\label{form:3}
\nabla_{\mu}\left(\frac{Rh\nabla^{\mu}f_1}{\sqrt{1+h\left(\nabla f_1\right)^2}}\right)=0\;,
\end{eqnarray}
 and
 \begin{eqnarray}\label{form:4}
\nabla_{\mu}\left(\sum_{i}\mathscr{L}^{i}\frac{h\nabla^{\mu}f_2}{\sqrt{1+h\left(\nabla f_2\right)^2}}\right)=0\;.
\end{eqnarray}
The equations of motion from Eq.~(\ref{form:1}) to  Eq.~(\ref{form:4}) are the main results of this study. We point out that there is no equation of motion for $h(x^{\alpha})$ since the action lacks the derivative of $h(x^{\alpha})$. Therefore $h(x^{\alpha})$ should be given in advance. In the next sections, we shall consider $h=1$ for simplicity.

\section{equations of motion in cosmology}
In this section, we present the equations of motion in the background of Friedman-Robertson-Walker universe
\begin{eqnarray}
ds^2=-dt^2+a\left(t\right)^2\left(dr^2+r^2 d\Omega^2\right)\;,
\end{eqnarray}
where $a(t)$ is the scale factor and $d\Omega^2$ is the line element of two dimensional unit sphere.
Except for perfect fluids, we also consider the quintessence field $\phi$ in the matter components with the Lagrangian density
\begin{eqnarray}
\mathscr{L}_{\phi}=\frac{1}{2}\left(\nabla\phi\right)^2+V\left(\phi\right)\;,
\end{eqnarray}
with $V(\phi)$ a scalar potential. We obtain the Einstein equations as follows
\begin{eqnarray}
&&3H^2-3 H\dot{f_1}\ddot{f_1}+3\left(\dot{H}+H^2\right)\dot{f_1}^2=8\pi\left[\left(\frac{1}{2}\dot{\phi}^2+V-\dot{\phi}^2\dot{f_2}^2+\sum_{i}\rho^{i}\right)
\sqrt{\frac{1-\dot{f_1}^2}{1-\dot{f_2}^2}}\right]\;,
\end{eqnarray}

\begin{eqnarray}
&&2\dot{H}+3H^2-\frac{\dot{f_1}^2\ddot{f_1}^2}{\left(1-\dot{f_1}^2\right)^2}
-\frac{\left(\ddot{f_1}^2+\dot{f_1}\dddot{f_1}+2H\dot{f_1}\ddot{f_1}\right)}{1-\dot{f_1}^2}=-8\pi\left[\left(\frac{1}{2}\dot{\phi}^2-V+\sum_{i}p^{i}\right)
\sqrt{\frac{1-\dot{f_2}^2}{1-\dot{f_1}^2}}
\right]\;,
\end{eqnarray}
The equation of motion for $\phi$ is
\begin{eqnarray}
\ddot{\phi}+3H\dot{\phi}+\frac{\partial V}{\partial\phi}-\frac{\dot{\phi}\dot{f_2}\ddot{f_2}}{1-\dot{f_2}^2}=0\;.
\end{eqnarray}
The equations determining the minimal surfaces are

\begin{eqnarray}
\left(\frac{a^3\dot{f_1}R}{\sqrt{1-\dot{f_1}^2}}\right)^{\cdot}=0\;,
\end{eqnarray}

and

\begin{eqnarray}
\left[\left(\frac{1}{2}\dot{\phi}^2-V+\sum_{i}\rho^{i}\right)\frac{ a^3\dot{f_2}}{\sqrt{1-\dot{f_2}^2}}\right]^{\cdot}=0\;.
\end{eqnarray}
where $H=\dot{a}/a$ is the Hubble parameter and dot denotes the derivative with respect to cosmic time $t$. Given the equation of state
\begin{eqnarray}
p^i=p^i\left(\rho^i\right)\;,
\end{eqnarray}
the system of equations (18-23) are closed. But they are rather involved. So we shall consider two cases. One is to assume only gravity on the dynamical minimal surface and the other is only quintessence on the dynamical surface.

\section{gravity on dynamical minimal surface}
In the absence of matters or only gravity on the dynamical surface, the Einstein equations become (for simplicity, we replace $f_1$ with $f$)
\begin{eqnarray}
&&3H^2-3 H\dot{f}\ddot{f}+3\left(\dot{H}+H^2\right)\dot{f}^2=0\;,
\end{eqnarray}
\begin{eqnarray}
&&2\dot{H}+3H^2-\frac{\dot{f}^2\ddot{f}^2}{\left(1-\dot{f}^2\right)^2}
-\frac{\left(\ddot{f}^2+\dot{f}\dddot{f}+2H\dot{f}\ddot{f}\right)}{1-\dot{f}^2}=0\;.
\end{eqnarray}
We can read the energy and pressure of dark energy
\begin{eqnarray}
&&\rho_{\Lambda}=\frac{3}{8\pi}\left[H\dot{f}\ddot{f}-\left(\dot{H}+H^2\right)\dot{f}^2\right]\;,
\end{eqnarray}
\begin{eqnarray}
&&p_\Lambda=-\frac{1}{8\pi}\left[\frac{\dot{f}^2\ddot{f}^2}{\left(1-\dot{f}^2\right)^2}
+\frac{\left(\ddot{f}^2+\dot{f}\dddot{f}+2H\dot{f}\ddot{f}\right)}{1-\dot{f}^2}\right]
\;.
\end{eqnarray}

The equation determining the minimal surface is
\begin{eqnarray}
\left(\frac{a^3\dot{f}R}{\sqrt{1-\dot{f}^2}}\right)^{\cdot}=0\;.
\end{eqnarray}
In order to study the dynamics of the system, we write the solution in the astomous form. To this end, we
define
\begin{eqnarray}
X\equiv\dot{f}\;,\ \ \ \ Y\equiv\frac{\ddot{f}}{H}\;,\ \ \ \
\end{eqnarray}
then the system of equations  are
\begin{eqnarray}
\frac{dX}{dN}&=&Y\;,\nonumber\\
\frac{dY}{dN}&=&\frac{X^4Y^2+3X^5Y+3XY-2X^2Y^2+5X^2-6X^3Y-4X^4-2+X^6}{X^3\left(1-X^2\right)}\;,
\end{eqnarray}
with
\begin{eqnarray}
N&=&\ln a\;.
\end{eqnarray}
We find there are three branches of solutions, solution I, solution II and solution III.  Solution I is given by
\begin{eqnarray}
a\propto\sqrt{t}\;,\ \ \ \ f=t\;.
\end{eqnarray}
This is a radiation dominated universe. For solution II and solution III , we plot the equation of state
\begin{eqnarray}
w\equiv\frac{p_{\Lambda}}{\rho_{\Lambda}}\;,
\end{eqnarray}
in Fig.~(3). It is found that Solution I, II and III have $w=1/3$,  $w>1/3$ and  $w<1/3$ respectively. In particular, solution III can cross the
phantom divide. Thus it is a potential candidate of dark energy.

\begin{figure}[h]
\includegraphics[width=8cm]{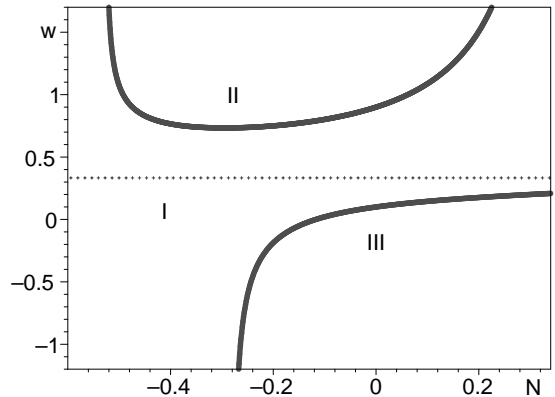}
\caption{The equation of state $w$ for three branches of solutions.}
\end{figure}

\section{quintessence on dynamical minimal surface}
In this section, we assume quintessence on the dynamical surface. We have the equations of motion as follows

\begin{eqnarray}\label{DE:1}
&&3H^2=8\pi\left[\left(\frac{1}{2}\dot{\phi}^2+V-\dot{\phi}^2\dot{f}^2\right)
\frac{1}{\sqrt{{1-\dot{f}^2}}}+\rho_m\right]\;,
\end{eqnarray}

\begin{eqnarray}\label{DE:2}
&&2\dot{H}+3H^2=-8\pi\left(\frac{1}{2}\dot{\phi}^2-V\right)
\sqrt{{1-\dot{f}^2}}\;,
\end{eqnarray}
where $\rho_m$ is the energy density of dark matter which has the equation of state $w=0$.
We conclude the density and pressure of dark energy
\begin{eqnarray}
&&\rho_{\Lambda}=\left(\frac{1}{2}\dot{\phi}^2+V-\dot{\phi}^2\dot{f}^2\right)
\frac{1}{\sqrt{{1-\dot{f}^2}}}\;,
\end{eqnarray}
\begin{eqnarray}
&&p_{\Lambda}=\left(\frac{1}{2}\dot{\phi}^2-V\right)
\sqrt{{1-\dot{f}^2}}\;.
\end{eqnarray}
The equation of motion for $\phi$ is
\begin{eqnarray}
\ddot{\phi}+3H\dot{\phi}+\frac{\partial V}{\partial\phi}-\frac{\dot{\phi}\dot{f}\ddot{f}}{1-\dot{f}^2}=0\;.
\end{eqnarray}
The equation determining the minimal surface is

\begin{eqnarray}
\left[\left(\frac{1}{2}\dot{\phi}^2-V\right)\frac{ a^3\dot{f}}{\sqrt{1-\dot{f}^2}}\right]^{\cdot}=0\;.
\end{eqnarray}
In order to make dynamical analysis of the system, we should write the equations of motion in the autonomous form.
So we define
\begin{eqnarray}
X\equiv\sqrt{\frac{8\pi}{3}}\frac{\dot{\phi}}{H}\;,\ \ \ \ Y\equiv\sqrt{\frac{8\pi}{3}}\frac{\sqrt{V}}{H}\;,\ \ \ \ Z\equiv\dot{f}\;,
\end{eqnarray}
then we obtain the system of equations
\begin{eqnarray}
\frac{dX}{dN}&=&\sqrt{3}\lambda Y^2\frac{-X^2+2Y^2+2X^2Z^2}{X^2-2Y^2+2X^2Z^2}-\frac{3}{2}X\left[1-\sqrt{1-Z^2}\left(\frac{X^2}{2}-Y^2\right)\right]+\frac{3XZ^2\left(X^2+2Y^2\right)}{X^2-2Y^2+2X^2Z^2}\;,\\
\frac{dY}{dN}&=&\frac{\sqrt{3}}{2}\lambda XY+\frac{3}{2}Y\left[1+\left(\frac{X^2}{2}-Y^2\right)\sqrt{1-Z^2}\right]\;,\\
\frac{dZ}{dN}&=&Z\left(1-Z^2\right)\frac{3X^2+4\sqrt{3}\lambda XY^2+6Y^2}{X^2-2Y^2+2X^2Z^2}\;,\\
\frac{d\lambda}{dN}&=&\sqrt{3}\lambda^2X\left(\Gamma-1\right)\;,
\end{eqnarray}
where
\begin{eqnarray}
\lambda\equiv\frac{V_{,\phi}}{\sqrt{8\pi}V}\;,\ \ \ \ \  \Gamma\equiv\frac{V_{,\phi\phi}V}{V_{,\phi}^2}.
\end{eqnarray}

\begin{table*}[t]
\begin{center}
\begin{tabular}{|c|c|c|c|c|c|c|c|}
\hline Name &  $X$ & $Y$ & $Z$ & Existence & Stability & $\Omega_\phi$
 & $w$ \\
\hline \hline (a) & 0 & 0 & 0 & All $\lambda$ & Unstable saddle
&   0 & -- \\
\hline \hline (b1) & $\sqrt{2}$ & 0 & 0 & All $\lambda$ & Unstable node for $\lambda\geq-\sqrt{6}$\;;
&   1 & 1 \\
& & & & & Unstable saddle for $\lambda<-\sqrt{6}$ & & \\
\hline \hline (b2) &$-\sqrt{2}$ & 0 & 0 & All $\lambda$ & Unstable node for $\lambda\leq\sqrt{6}$\;;
&   1 & 1 \\
& & & & & Unstable saddle for $\lambda>\sqrt{6}$ & & \\
\hline \hline (c) & $-\frac{\sqrt{3}\lambda}{3}$ & $\sqrt{1-\frac{\lambda^2}{6}}$ & 0 & All $\lambda$ & Stable node for $\lambda^2\leq3$\;;
 &   1 & $\frac{\lambda^2}{3}-1$ \\
& & & & & Unstable node for $\lambda^2\geq 6$\;; & & \\
& & & & & Unstable saddle for $3<\lambda^2<6$ & & \\
\hline \hline (d) & $-\frac{\sqrt{3}}{\lambda}$ & $\frac{\sqrt{6}}{2\lambda}$ & Z & All $\lambda$& Stable line segment
& $\frac{3}{\lambda^2}\sqrt{1-Z^2}$ & 0 \\
\hline
\end{tabular}
\end{center}
\caption[crit]{The properties of the critical points for the exponential potential given by
$V\propto e^{\sqrt{8\pi}\lambda\phi}$. } \label{crit0}
\end{table*}

\begin{figure}[h]
\includegraphics[width=8cm]{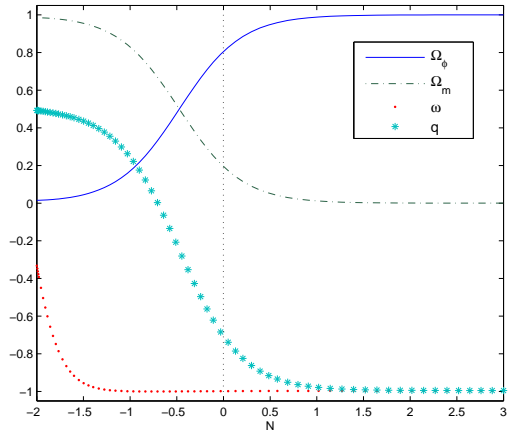}
\caption{The evolution of $\Omega_{\phi}, \Omega_{m}, w, q$ with the scale factor $N\equiv\ln a$ for exponential potential $V\propto e^{\sqrt{8\pi}\lambda\phi}$.}
\end{figure}

We have studied the exponential potential and the power-law potential,
\begin{eqnarray}
V\left(\phi\right)\propto e^{\sqrt{8\pi}\lambda\phi}\;,\ \ \ \ V\left(\phi\right)\propto \phi^n\;,
\end{eqnarray}
 respectively. For the exponential potential, $\lambda$ is a constant and we are left with three dynamical equations. In Table I, we present the critical points of the system. It is found there are five critical points. Point (a) corresponds to the matter dominated epoch of the universe. It is an unstable saddle point.  Points (b) corresponds to the quintessence dominated epoch. They are either unstable node or unstable saddle points dependant on the parameter, $\lambda$.  Point (c) corresponds to the quintessence dominated epoch. But it can be unstable node, unstable saddle and stable node dependant on $\lambda$. The existence of stable node is important because it reveals the universe is doomed to evolve into the dark energy dominated epoch. If $\lambda^2\ll 3$, we have the equation of state for dark energy $w\simeq -1$. Point (d) corresponds to the epoch of matter and quintessence co-dominant. In this case, quintessence has the same equation of sate with matter. Finally, point (d) is actually a stable line segment. In Fig.~(5), by putting $\lambda=0.1$, we plot the evolution of equation of state $w$, the decelerating parameter $q$ and the ratios $\Omega_{\Lambda}=\frac{8\pi\rho_{\Lambda}}{3H^2}$ for dark energy and $\Omega_m=\frac{8\pi\rho_m}{3H^2}$ for dark matter. They are consistent with the observations very well.

For the power-law potential, $\lambda$ is not a constant but $\Gamma=\frac{n-1}{n}$ is and we are left with four dynamical equations. In Table II, we present the critical points of the system. In this case, there are six critical points. Points (a) and ($c_1, c_2$)  correspond to the matter dominated epoch. They are all unstable saddle points.  Points ($d_1, d_2$) corresponds to the quintessence dominated epoch but they all have the equation of state $w=+1$. This means quintessence ever dominates the universe before the dark matter (and radiation) dominated epoch. They are unstable nodes. Finally, point ($b$) represents the dark energy dominated universe. It is a stable node and has the equation of state exactly $w=-1$ regardless of the parameter $\Gamma$. In Fig.~(6) and Fig.~(7), by putting $n=2$, we plot the evolution of equation of state $w$, the decelerating parameter $q$ and the ratios $\Omega_{\Lambda}$,  $\Omega_m$ for dark energy and dark matter. They are consistent with the observations.

\begin{table*}[t]
\begin{center}
\begin{tabular}{|c|c|c|c|c|c|c|c|c|}
\hline Name &  $X$ & $Y$ & $Z$ & $\lambda$& Existence & Stability & $\Omega_\phi$
 & $w_\phi$ \\
\hline \hline (a) & 0 & 0 & 0 & $\lambda$ & All $\Gamma$ & Unstable saddle
&   0 & --  \\
\hline \hline (b) &0 &1 & 0 & 0 & All $\Gamma$ & Stable node
&   1 & -1 \\
\hline \hline (c1) & 0 & 0 &1  &$\lambda$ & All $\Gamma$ &  Unstable saddle
 &  0 & -- \\
\hline \hline (c2) & 0& 0 & -1& $\lambda$ & All $\Gamma$ & Unstable saddle
 &   0 & --\\
\hline \hline (d1) & $\sqrt{2}$ & 0 &0 & 0 & All $\Gamma$&  Unstable node
&  1 & 1 \\
\hline \hline (d2) & $-\sqrt{2}$ & 0& 0 &0 & All $\Gamma$&  Unstable node
& 1 & 1\\
\hline
\end{tabular}
\end{center}
\caption[crit]{The properties of the critical points for the power law potential given by
$V\propto \phi^{2}$. } \label{crit0}
\end{table*}

\begin{figure}[h]
\includegraphics[width=8cm]{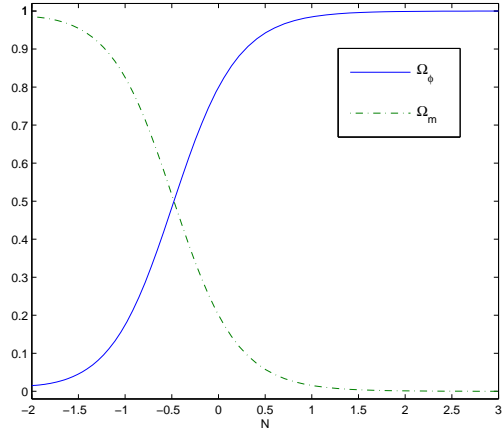}
\caption{The evolution of dark energy and dark matter ratios  for the power-law potential $V\propto \phi^{2}$.}
\end{figure}

\begin{figure}[h]
\includegraphics[width=8cm]{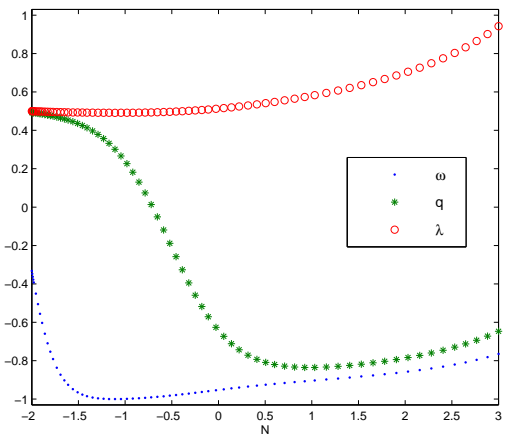}
\caption{The evolution of $w,q,\lambda$ for power-law potential $V\propto \phi^{2}$.}
\end{figure}

\section{Conclusion and Discussion}\label{sec:8}
String theory reveals that the spacetime must be higher than four on small scales. Further more, the brane world models
tell us the extra dimensions not only exist bout also can be very large (or even infinite). On the other hand,  minimal surfaces have many wonderful properties such as minimum area, minimum energy and so on. Motivated by these ideas, we propose our 4-dimensional universe locates on a minimal surface embedded in a 5-dimensional spacetime. Concretely, we construct the equations of motion when gravity and matters are confined on different minimal surfaces. In the background of Friedman-Robertson Walker universe, we derive the dynamical equations in the presence of matters and quintessence. As one example, we investible the comic evolution when gravity is confined on the minimal surface. We find the corresponding equation of state for dark energy varies from $-\infty$ to $+\infty$. This enables it to be a potential candidate of dark energy. As the second example we investigate the case when quintessence confined on the minimal surface. We find there are always stable nodes whether for exponential potential or power-law potential. This makes quintessence to be a very good candidate for dark energy when it is confined on the minimal surface.

Finally, from the equations of motion, we conclude the physics of both gravity and matters may be significantly influenced due to the presence of minimal surfaces.
Thus a number of questions as well as more
fundamental issues, such as the Newtonian and post-Newtonian limits, the cosmic perturbations, the weak and strong self-field effects, the solutions for stars and black holes et. al remain open and deserve to be studied.

\section*{Acknowledgments}
This work is partially supported by China Program of International ST Cooperation 2016YFE0100300
, the Strategic Priority Research Program ``Multi-wavelength Gravitational Wave Universe'' of the
CAS, Grant No. XDB23040100, the Joint Research Fund in Astronomy (U1631118), and the NSFC
under grants 11473044, 11633004 and the Project of CAS, QYZDJ-SSW-SLH017.

\newcommand\ARNPS[3]{~Ann. Rev. Nucl. Part. Sci.{\bf ~#1}, #2~ (#3)}
\newcommand\AL[3]{~Astron. Lett.{\bf ~#1}, #2~ (#3)}
\newcommand\AP[3]{~Astropart. Phys.{\bf ~#1}, #2~ (#3)}
\newcommand\AJ[3]{~Astron. J.{\bf ~#1}, #2~(#3)}
\newcommand\GC[3]{~Grav. Cosmol.{\bf ~#1}, #2~(#3)}
\newcommand\APJ[3]{~Astrophys. J.{\bf ~#1}, #2~ (#3)}
\newcommand\APJL[3]{~Astrophys. J. Lett. {\bf ~#1}, L#2~(#3)}
\newcommand\APJS[3]{~Astrophys. J. Suppl. Ser.{\bf ~#1}, #2~(#3)}
\newcommand\JHEP[3]{~JHEP.{\bf ~#1}, #2~(#3)}
\newcommand\JMP[3]{~J. Math. Phys. {\bf ~#1}, #2~(#3)}
\newcommand\JCAP[3]{~JCAP {\bf ~#1}, #2~ (#3)}
\newcommand\LRR[3]{~Living Rev. Relativity. {\bf ~#1}, #2~ (#3)}
\newcommand\MNRAS[3]{~Mon. Not. R. Astron. Soc.{\bf ~#1}, #2~(#3)}
\newcommand\MNRASL[3]{~Mon. Not. R. Astron. Soc.{\bf ~#1}, L#2~(#3)}
\newcommand\NPB[3]{~Nucl. Phys. B{\bf ~#1}, #2~(#3)}
\newcommand\CMP[3]{~Comm. Math. Phys.{\bf ~#1}, #2~(#3)}
\newcommand\CQG[3]{~Class. Quantum Grav.{\bf ~#1}, #2~(#3)}
\newcommand\PLB[3]{~Phys. Lett. B{\bf ~#1}, #2~(#3)}
\newcommand\PRL[3]{~Phys. Rev. Lett.{\bf ~#1}, #2~(#3)}
\newcommand\PR[3]{~Phys. Rep.{\bf ~#1}, #2~(#3)}
\newcommand\PRd[3]{~Phys. Rev.{\bf ~#1}, #2~(#3)}
\newcommand\PRD[3]{~Phys. Rev. D{\bf ~#1}, #2~(#3)}
\newcommand\RMP[3]{~Rev. Mod. Phys.{\bf ~#1}, #2~(#3)}
\newcommand\SJNP[3]{~Sov. J. Nucl. Phys.{\bf ~#1}, #2~(#3)}
\newcommand\ZPC[3]{~Z. Phys. C{\bf ~#1}, #2~(#3)}
\newcommand\IJGMP[3]{~Int. J. Geom. Meth. Mod. Phys.{\bf ~#1}, #2~(#3)}
\newcommand\IJMPD[3]{~Int. J. Mod. Phys. D{\bf ~#1}, #2~(#3)}
\newcommand\IJMPA[3]{~Int. J. Mod. Phys. A{\bf ~#1}, #2~(#3)}
\newcommand\GRG[3]{~Gen. Rel. Grav.{\bf ~#1}, #2~(#3)}
\newcommand\EPJC[3]{~Eur. Phys. J. C{\bf ~#1}, #2~(#3)}
\newcommand\PRSLA[3]{~Proc. Roy. Soc. Lond. A {\bf ~#1}, #2~(#3)}
\newcommand\AHEP[3]{~Adv. High Energy Phys.{\bf ~#1}, #2~(#3)}
\newcommand\Pramana[3]{~Pramana.{\bf ~#1}, #2~(#3)}
\newcommand\PTP[3]{~Prog. Theor. Phys{\bf ~#1}, #2~(#3)}
\newcommand\APPS[3]{~Acta Phys. Polon. Supp.{\bf ~#1}, #2~(#3)}
\newcommand\ANP[3]{~Annals Phys.{\bf ~#1}, #2~(#3)}
\newcommand\RPP[3]{~Rept. Prog. Phys. {\bf ~#1}, #2~(#3)}

\end{document}